\documentclass[aps,pra,epsfigure,twocolumn,superscriptaddress]{revtex4-1}
\usepackage[colorlinks=true,linkcolor=blue,urlcolor=blue,citecolor=blue,pdfusetitle]{hyperref}
\usepackage[utf8]{inputenc}
\usepackage[english]{babel}
\usepackage{amsmath}
\usepackage[caption = false]{subfig}
\usepackage{graphicx,epstopdf}
\usepackage{blindtext}
\usepackage[table,xcdraw]{xcolor}
\usepackage{lipsum}
\usepackage{amsfonts}
\usepackage{bbm}
\usepackage{amssymb}
\usepackage{enumerate}
\usepackage{color}
\usepackage{latexsym}
\usepackage{physics}
\usepackage{times,txfonts}
\newtheorem{theorem}{Theorem}

\newtheorem{condition}[theorem]{Condition}

\newcommand{\Dcal}{\mathcal{D}}
\newcommand{\Ecal}{\mathcal{E}}

\newcommand{\Gcal}{\mathcal{G}}
\newcommand{\Hcal}{\mathcal{H}}
\newcommand{\Ical}{\mathcal{I}}

\newcommand{\Lcal}{\mathcal{L}}

\newcommand{\Ocal}{\mathcal{O}}

\newcommand{\Rcal}{\mathcal{R}}

\newcommand{\Tcal}{\mathcal{T}}

\newcommand{\1}{\mathbbm{1}}
\newcommand{\Lmath}{\mathbbm{L}}
\newcommand{\Rmath}{\mathbbm{R}}

\newcommand{\dket}[1]{| #1 \rangle\rangle}
\newcommand{\dbra}[1]{\langle\langle #1 |}

\newcommand{\dinterpro}[2]{\langle \langle #1 | #2 \rangle \rangle}

\newcommand{\trs}[1]{\text{Tr}(#1)}

\newcommand{\cvb}[1]{{\color{blue}#1}}
\definecolor{greendark}{rgb}{0.0, 0.5, 0.0}

\begin{document}
	
	\title{Quantum adiabatic brachistochrone for open systems}
	
	\author{Alan C. Santos}
	\email{ac\_santos@df.ufscar.br}
	\affiliation{Departamento de F\'{i}sica, Universidade Federal de S\~ao Carlos, P.O. Box 676, 13565-905, São Carlos, São Paulo, Brazil}
	
	\author{C. J. Villas-Boas} 
	\email{celsovb@df.ufscar.br}
	\affiliation{Departamento de F\'{i}sica, Universidade Federal de S\~ao Carlos, P.O. Box 676, 13565-905, São Carlos, São Paulo, Brazil}
	
	\author{R. Bachelard} 
	\email{bachelard.romain@gmail.com}
	\affiliation{Departamento de F\'{i}sica, Universidade Federal de S\~ao Carlos, P.O. Box 676, 13565-905, São Carlos, São Paulo, Brazil}
	
	\begin{abstract}
		We propose a variational principle to compute a quantum adiabatic brachistochrone (QAB) for open systems. Using the notion of ``adiabatic speed" based on the energy gaps, we derive a Lagrangian associated to the functional measuring the time spent to achieve adiabatic behavior, which in turn allows us to perform the optimization. The QAB is illustrated for non-unitary dynamics of STIRAP process, the Deutsch-Jozsa quantum computing algorithm and of a transmon qutrit. A numerical protocol is devised, which allows to compute the QAB for arbitrary quantum systems for which exact simulations can be afforded. We also establish sufficient conditions for the equivalence between the Lagrangians, and thus the QAB, of open and closed systems.
	\end{abstract}
	
	\maketitle
	
	\section{Introduction}
	
	The brachistochrone is a fundamental problem in classical mechanics~\cite{Goldstein:Book}, which aims at identifying the optimal trajectory that minimizes the travel time between two points for a particle submitted to gravity. This optimization problem has been extended to the quantum realm, where a \textit{quantum brachistochrone} is the time-optimal (unitary) path connecting two points of the Hilbert space~\cite{Carlini:06}. This quantum formulation of a variational principle to find the optimal dynamics has been used in different contexts~\cite{Borras:07,Borras:08,Frydryszak:08}, such as quantum computation~\cite{Kuzmak:13,Russell:15,Brody:15} and optimal control~\cite{Wang:15-2}. It is particularly promising in the field of quantum computation, where the notion of quantum adiabatic brachistochrone (QAB) for unitary dynamics has been introduced one decade ago by Rezakhani \textit{et al}~\cite{Rezakhani:09}. It allows one to efficiently implement quantum tasks using adiabatic dynamics~\cite{Garnerone:12,Marcela:14}.
	
	Despite the efforts to isolate quantum systems, it is necessary to take into account the inevitable influence of their environment on the dynamics. Differently from the ideal (closed) case where the dynamics is governed only by the Hamiltonian, a real quantum process is also driven by an additional set of parameters associated to the interaction with the environment~\cite{Lindblad:76}. 
	Several physical effects may stem from it, but the adiabatic dynamics is particularly affected, as the perfectly uncoupled dynamics of the Hilbert-Schrödinger eigenspaces start exchanging energy. A turnaround consists in casting the system as an independent dynamics of Lindblad-Jordan eigenspaces~\cite{Sarandy:05-1}, a formulation of adiabaticity  for open systems which has been applied to quantum computation~\cite{Sarandy:05-2}, state engineering~\cite{Hu:19-a,Jing:16} and quantum thermodynamics~\cite{Hu:19-d}, for example. In this context, finding new strategies do tackle optimization problems with adiabatic dynamics in presence of decoherence is a fundamental issue.
	
	In this work we propose a variational formulation for the QAB of open systems. To this end, the definition of ``adiabatic speed" is generalized, where the influence of relative quantal phases for the dynamics naturally emerges, as supported by recent experiments~\cite{Hu:19-a,Hu:19-d}. Effect of decoherence on the QAB is discussed for a single qubit and for a STIRAP process, which leads us to define criteria for the equivalence between the Lagrangians of closed and open systems. The gain provided by the brachistochrone trajectory is illustrated as a reduction in the time necessary to reach the target state, up to a given fidelity. Finally, we compute the QAB for the Deutsch-Jozsa algorithm with arbitrary particle number and under dephasing, discussing the potential implications for quantum computation.
	
	
	\section{Adiabatic brachistochrone for open systems}
	
	When the inevitable coupling of a quantum system with its environment is taken into account, one describes its state with a reduced density matrix $\rho(t)$, whose evolution is obtained from the open system master equation: $\dot{\rho}(t) = \Lcal[\rho(t)]$, 
	with $\Lcal[\bullet]=\!(1/i\hbar)[H(t),\bullet] + \Rcal[\bullet]$ the generator of the dynamics, which encodes the contribution of the Hamiltonian $H(t)$ of the system and the interaction between the system and its reservoir (environment), here described by the superoperator $\Rcal[\bullet]$. Then, the adiabatic behavior of an open system is defined from the spectrum of the superoperator $\Lmath(t)$ used to rewrite the system dynamics~(see Appendix~\ref{derivEq1}):
	\begin{eqnarray}
	\dket{\dot{\rho}(t)} = \Lmath(t)\dket{\rho(t)}. \label{DynSuper}
	\end{eqnarray}
	$\Lmath(t)$ here refers to the super-matrix (Lindblad superoperator) with matrix elements $\Lmath_{kl}(t)\!=\!\trs{\sigma_{k}^{\dagger} \Lcal [ \sigma_{l} ]}$, and $\dket{\rho(t)}$ to the ``coherence" vector with components $\varrho_{n}(t)\!=\!\trs{\rho(t)\sigma_{n}^{\dagger}}$, with $\sigma_{n}$ an element of the set of $D^2\!-\!1$ operators $\{\sigma_{n}\} \in \Hcal_{\text{S}}$ and $D$ the dimension of the Hilbert space $\Hcal_{\text{S}}$. Thus, one has $\trs{\sigma_{n}}\!=\!0$ and $\trs{\sigma_{n}\sigma_{m}^{\dagger}}\!=\!D\delta_{nm}$. Given that the superoperator $\Lmath(t)$ is not Hermitian, in general it does not admit a diagonal form, so that the adiabaticity definition is considered from its block Jordan form~\cite{Sarandy:05-1,Horn:Book}. The Jordan decomposition of the superoperator $\Lmath(s)$ is obtained from its left and right ``quasi"-eigenstates $\dbra{\Ecal_{\alpha}^{n_{\alpha}}(t)}$ and $\dket{\Dcal_{\alpha}^{n_{\alpha}}(t)}$: They satisfy $\Lmath(t)|\Dcal_{\alpha}^{n_{\alpha}}(t) \rangle\rangle \!=\!\lambda_{\alpha}(t)\dket{\Dcal_{\alpha}^{n_{\alpha}}(t)}+\dket{\Dcal_{\alpha}^{n_{\alpha}-1}(t)}$ and $\dbra{\Ecal_{\alpha}^{n_{\alpha}}(t)}\Lmath(t)\!=\!\lambda_{\alpha}(t)\dbra{\Ecal_{\alpha}^{n_{\alpha}}(t)}+\dbra{\Ecal_{\alpha}^{n_{\alpha}+1}(t)}$, respectively, where $n_{\alpha}$ denotes the $n$th quasi-eigenvector of $\Lmath(t)$ with eigenvalue $\lambda_{\alpha}(t)$, and $N_{\alpha}$ hereafter refers to the dimension of the $\alpha$-th block~\cite{Sarandy:05-1,Horn:Book}. 
	
	The extension of adiabaticity to open system is provided by the independent evolution of different Jordan blocks with distinct and non-crossing instantaneous eigenvalues $\lambda_{\alpha}(t)$ of $\Lmath(t)$~\cite{Sarandy:05-1}. From this definition, it is possible to show that a \textit{sufficient} condition for adiabaticity is given by:
	\begin{gather}
	\left\vert \sum_{n_{\alpha}}^{N_{\alpha}}e^{\int_{0}^{t}\Re[\Gcal_{\alpha\beta}(\xi)]d\xi} \dbra{\Ecal_{\beta}^{n_{\beta}}(t)}\dot{\Lmath}(t)\dket{\Dcal_{\alpha}^{n_{\alpha}}(t)}\right \vert/|\Gcal_{\alpha\beta}(s)|^2\!\ll\!\epsilon
	\end{gather}
	for any $\alpha$ and $\beta$, and $\epsilon$ chosen to be arbitrarily small. Finally,
	$\Gcal_{\alpha\beta}(t)=\lambda_{\alpha}(t) - \lambda_{\beta}(t)$ is the instantaneous gap~\cite{Sarandy:05-1,Hu:19-a}.
	
	\emph{Lagrangian formalism --} The QAB problem for open systems relies on the definition of a Lagrangian, as for the case of closed systems~\cite{Rezakhani:09}. However, the definition of ``adiabatic speed" used for closed systems is no longer valid, which leads us to introducing the following generalized version:
	\begin{eqnarray}
	v_{\text{ad}}^{\text{os}}(s) = \frac{\epsilon |\Gcal(s)|^2}{\vert\vert\Lmath^{\prime}(s)\vert\vert }\exp(-\int_{0}^{t}\Re[\Gcal(\xi)]d\xi) , \label{vad}
	\end{eqnarray}
	with $s(t)$ the normalized time ($f^{\prime}(s)\!=\!df(s)/ds$), $\Gcal(s)$ the \textit{minimum non-vanishing} instantaneous gap, and $||A||^2\!=\!\trs{A^{\dagger}A}$. The exponential term in Eq.~\eqref{vad} enforces the adiabaticity condition for open systems~\cite{Sarandy:05-2,Hu:19-a}. The above definition is consistent with the definition of ``adiabatic speed" in closed system, since in that limit $\Rcal[\rho(t)]\!=\!0$ and thus $|\Gcal_{\alpha\beta}(s)|\!\propto\!|\Delta(s)|$, $\Re[\Gcal_{\alpha\beta}(t)]\!=\!0$ and $||\Lmath^{\prime}(s)||\!\propto\! ||H^{\prime}(s)||$, recovering the result discussed in Ref.~\cite{Rezakhani:09}. Therefore, the functional time is defined as $\Tcal_{\text{os}}\!=\!\int_{0}^{1} ds / v_{\text{ad}}^{\text{os}}(s)$, from which the Lagrangian for the brachistochrone problem is derived:
	\begin{eqnarray}\label{FunctionalOS}
	L_{\text{os}}[q^{\prime}(s),q(s)] = \frac{||\Lmath^{\prime}(s)||}{|\Gcal(s)|^2}\exp(\frac{1}{\tau}\int_{0}^{s}\Re[\Gcal(\xi)]d\xi)\text{ , }
	\end{eqnarray}
	with $q(s)$ the generalized parameter associated to the Hamiltonian and to the decoherence, and $\tau$ the total evolution time. As we shall see, there are several Lagrangians which provide the same solution to the Euler-Lagrange equation, and thus the same dynamics. For convenience, we consider a reparametrization of the Lagrangian as $\tilde{L}_{\text{os}}[q^{\prime}(s),q(s)]\!=\!L_{\text{os}}^2[q^{\prime}(s),q(s)]$ to be used for the optimization throughout this work. This operation preserves the length of the curve provided by the Euler-Lagrange invariant, as discussed in~\cite{Nielsen:06}, but it also presents the advantage to make our approach fully consistent with that for closed systems~\cite{Rezakhani:09}, in the limit of zero decoherence. 
	
	An important issue in our approach is the complexity of computing the above Lagrangian. $\Lmath(s)$ presents a quadratic growth with the dimension $d$ of the system Hilbert space, so that there is no guarantee that an analytical expression for $L_{\text{os}}$ can be derived for $d\!\geq\!3$; in turn, the absence of an explicit form for $L_{\text{os}}$ will prevent one from  deriving the QAB analytically. However, one can compute the brachistochrone numerically within the usual limitations of exact simulations, which involves computing the eigenvalues of $\Lmath$. Differently, the case of a qubit ($d\!=\!2$) always yields an analytical solution since the problem of finding the spectrum of $\Lmath(s)$ is equivalent to solving a fourth-order polynomial function, which can be done by the discriminant method as established by the Abel-Ruffini theorem~\cite{Ruffini:Book,Abel:12}. Finally, we remark that since the Lagrangian for Euler-Lagrange equation is invariant under multiplication of $|\Gcal(s)|$ by a constant factor, a dedicated analysis of the relevant minimum gap must be performed in the case of high-dimensional superoperator $\Lmath(s)$.
	
	\section{Applications}
	
	\subsection{Single-qubit dynamics under dephasing}
	
	As first example, let us consider the adiabatic dynamics of a driven qubit, with Hamiltonian
	\begin{equation}
	H(s)=\frac{\hbar}{2} \left[ \Omega_{x}(s) \sigma_{x} + \Omega_{y}(s) \sigma_{y} \right],
	\end{equation}
	with $\Omega_{x}(0)\!=\!\Omega_{y}(1)\!=\!\Omega_0$ and $\Omega_{x}(1)\!=\!\Omega_{y}(0)\!=\!0$ as boundary conditions, with $\sigma_{\alpha}$ ($\alpha =x,y,z$) the Pauli matrices. Such dynamics can be experimentally implemented in nuclear magnetic resonance setups~\cite{Peterson:18}, for example. We consider the system-reservoir interaction as $\Rcal[\bullet]\!=\!\gamma(\sigma_{z}\bullet\sigma_{z}-\bullet)$, describing a dephasing channel. In the basis 
	$\sigma_{\text{tls}}\!=\!\{ \1,\sigma_{x},\sigma_{y},\sigma_{z}\}$, the superoperator $\Lmath(s)$ reads:
	\begin{eqnarray}
	\Lmath(s) = \begin{bmatrix}
	0 & 0 & 0 & 0 \\
	0 & -2\gamma & 0 & \Omega_{y}(s)  \\
	0 & 0 & -2\gamma & -\Omega_{x}(s) \\
	0 & -\Omega_{y}(s) & \Omega_{x}(s) & 0 
	\end{bmatrix} \text{ . }
	\end{eqnarray}
	The eigenvalues of $\Lmath(s)$ are $\lambda_{0}\!=\!0$, $\lambda_{1}\!=\!-2\gamma$ and $\lambda_{\pm}(s)\!=\!-\gamma \pm i \Delta(s)$, with $ \Delta^2(s) \!=\! \Omega^2_{x}(s)+\Omega^2_{y}(s) -\gamma^2$. From this set of eigenvalues we identify the minimum non-vanishing gap as $|\tilde{\Gcal}_{\pm\mp}(s)|\!=\!2|\Delta(s)|$~\footnote{In fact, different than other gaps, it is possible to show that $|\tilde{\Gcal}_{\pm\mp}(s)|\!\rightarrow\!0$ if, and only if, $\gamma\!\rightarrow\!0$ and $\Omega_{0}\!\rightarrow\!0$, simultaneously, so that the system does not evolve at all.}. Therefore, the Lagrangian for our system is given, up to a constant factor, by:
	\begin{eqnarray}
	\tilde{L}_{\text{os}}[\Omega(s),\Omega^{\prime}(s)] = \frac{ [\Omega^{\prime}_{x}(s)]^2 + [\Omega^{\prime}_{y}(s)]^2}{\left( \Omega^2_{x}(s)+\Omega^2_{y}(s) -\gamma^2\right)^2}  \text{ . } \label{LtlsOS}
	\end{eqnarray}
	\begin{figure}[t!]
		\centering
		\subfloat[ ]{\includegraphics[scale=0.2]{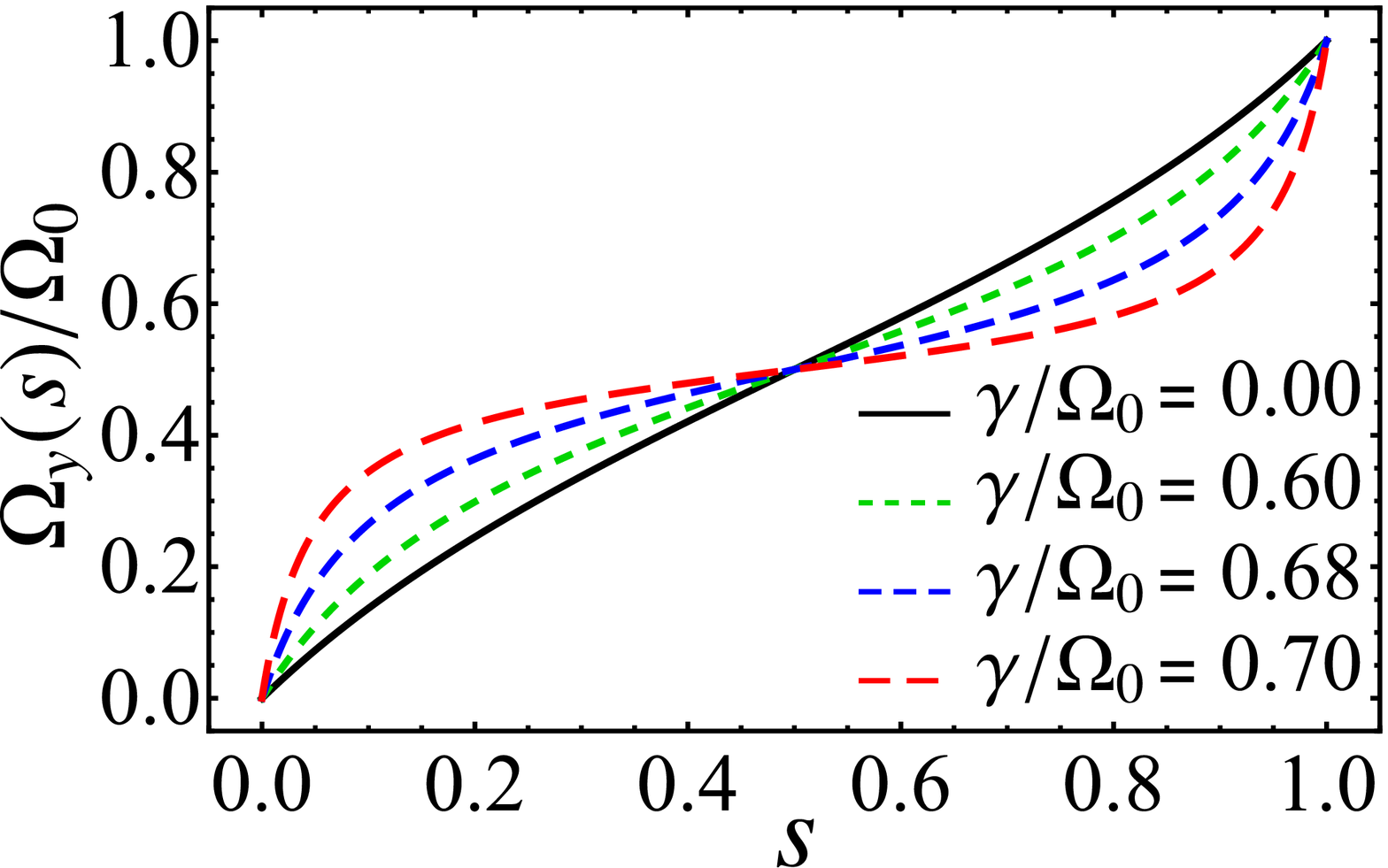}\label{Fig1a}}~
		\subfloat[ ]{\includegraphics[scale=0.2]{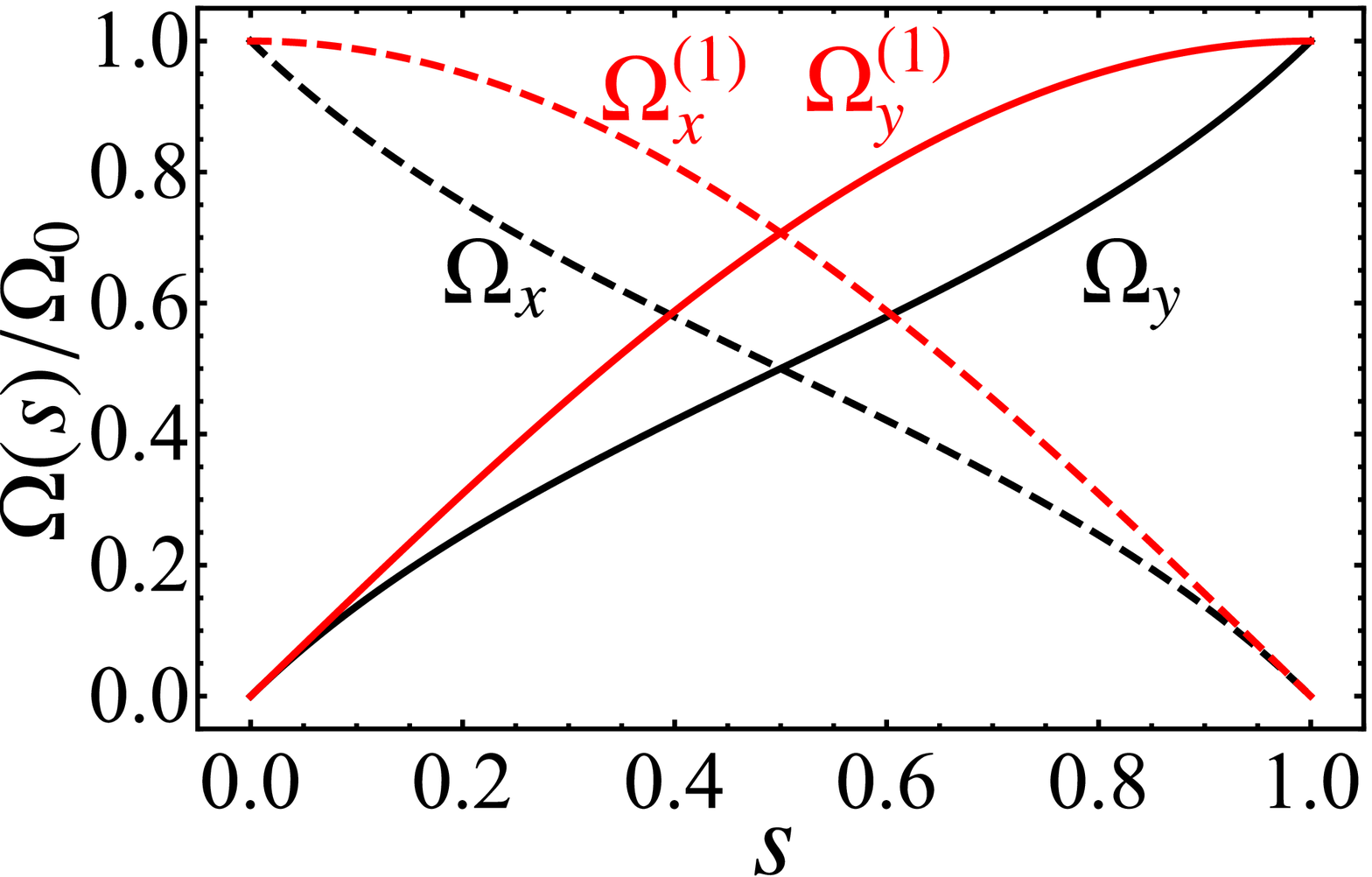}\label{Fig1b}}
		\caption{\eqref{Fig1a} Brachistochrone curve $\Omega_y(s)$ (with $\Omega_x(s)=1-\Omega_y(s)$) for an increasing decoherence rate $\gamma$. \eqref{Fig1b} Brachistochrone curves from Eqs.~\eqref{TLSbrachi} and~\eqref{LtlsOS2},  respectively, for the case $\gamma/\Omega\!=\!0.1$.}\label{Fig1}
	\end{figure}
	From this Lagrangian one can derive the optimal trajectory by employing the Euler-Lagrange equations. However, this is not a simple task for the general case, since it requires the solution of two coupled second-order differential equations, due to the presence of two free parameters $\Omega_x(s)$ and $\Omega_y(s)$. In order to circumvent this problem and provide analytical examples, we consider some specific constraints.
	As a first example, we impose $\Omega_x(s)+\Omega_y(s)\!=\!\Omega_0$, leading to:
	\begin{eqnarray}
	\Omega_{y}(s) = \frac{\Omega_0}{2}-\frac{\tilde{\Omega}_{0}}{2}  \tan \left( (1-2s) \arctan [\Omega_{0}/\tilde{\Omega}_{0}]\right)\text{ , } \label{TLSbrachi}
	\end{eqnarray}
	with $\tilde{\Omega}_{0}^2\!=\!|\Omega_{0}^2 - 2 \gamma^2|$. In the limit of zero decoherence ($\gamma\!\rightarrow\!0$), the above solution reduces to the QAB obtained in~\cite{Rezakhani:09}. The effect of the decoherence on the brachistrochrone is presented in Fig.~\ref{Fig1a}, which can be explained as follows: As can be observed from Eq.~\eqref{TLSbrachi}, the maximum gap $|\Gcal(s)|$ occurs at the initial and final time ($s\!=\!0$ and $1$), while the minimum one is found at $s\!=\!1/2$. Since in the present case the ``adiabatic speed" scales as $v_{\text{ad}}^{\text{os}}(s)\!\propto\!|\Gcal(s)|^2 / |\Omega_{y}^{\prime}(s)|^2$, preserving a finite speed leads to a brachistochrone with a sharp variation in $\Omega_y$ at initial and final time, yet a slow one around $s\!=\!1/2$. Finally, the larger the decoherence rate $\gamma$, the smaller the minimum gap at $s\!=\!1/2$, and the sharper the variations in the QAB curve. 
	
	As another possible constraint, we consider $\Omega_{x}^{2}(s) + \Omega_{y}^{2}(s)\!=\!\Omega_0^{2}$, which can be interpreted as a constant total power, since the Rabi frequencies $\Omega_{x,y}$ are proportional to the field amplitudes~\cite{Peterson:18}.  Then, the optimal fields as predicted by the brachistrochrone read
	\begin{eqnarray}
	\Omega_{x}^{(1)}(s) = \Omega_0\cos(\pi s/2) \text{ , } ~~ \Omega_{y}^{(1)}(s) = \Omega_0\sin(\pi s/2) \text{ . } \label{LtlsOS2}
	\end{eqnarray}
	The pump profiles for these two constraints are presented in Fig.\ref{Fig1b}, where one can observe that the QAB associated to Eq.~\eqref{LtlsOS} requires overall less energy than that of Eq.~\eqref{LtlsOS2}: It reflects that the former case is characterized by a smaller energy gap, and thus a longer time to achieve adiabaticity, than the latter~\cite{Sarandy:04,Amin:09}.
	We also remark that the dependence on $\gamma$ has now vanished, i.e., the brachistrochrone is the same for the open system whatever the dephasing rate as for the closed one. As we shall now discuss, this peculiarity originates in the equivalence between the Lagrangians for the closed and open systems. 

	
	\subsection{Equivalent Lagrangians}
	
	As it appears from the above example, the presence of a dephasing may not alter the brachistochrone curve. This leads us to establish a \textit{sufficient condition} for the QAB to be the same with and without decoherence.
	First, we consider two Lagrangians $L_{1}[q,q^{\prime}]$ and $L_{2}[q,q^{\prime}]$ to be equivalent if there exists a constant $c$ so that $L_{1}[q,q^{\prime}]\!=\!c L_{2}[q,q^{\prime}]$.
	Then we can show that:
	\begin{condition}\label{Cond2}
		A open system Lagrangian $L_{\mathrm{os}}[q,q^{\prime}]$ is equivalent to a closed system one $L_{\mathrm{cs}}[q,q^{\prime}]$ whenever the decoherence rates are constants and the minimum gap is a time-independent purely imaginary complex number.
	\end{condition}
	The proof follows from two observations: First, one can write the master equation in the form $\Lcal[\bullet]\!=\! [H(s),\bullet] + \Rcal_{t}[\bullet]$, with $\Rcal_{t}[\bullet]$ denoting the interaction with a time-dependent reservoir. For time-independent decoherence rates we can write the elements of $\Lmath(s)$ as $\Lmath_{kl}(s)\!=\!\trs{\sigma_{k}^{\dagger} [H(s),\sigma_{l} ]}/i\hbar + \trs{\sigma_{k}^{\dagger} \Rcal [ \sigma_{l} ]}$, 
	making clear that $\Lmath_{kl}^{\prime}(s)\!=\!\trs{\sigma_{k}^{\dagger} [H^{\prime}(s),\sigma_{l} ]}/i\hbar$, that is, $\Lmath_{kl}^{\prime}(s)$ is equivalent to the Lindblad superoperator for closed systems. Therefore, there is a number $c_{0}$ such that $||\Lmath^{\prime}(s)||\!=\!c_{0}||H^{\prime}(s)||$. Second, since the open system minimum gap is time-independent $\Gcal_{\text{os}}(s)\!=\!\Gcal_{\text{os}}$, then the minimum gap in the closed system is also time-independent, $g_{\text{cs}}(s)\!=\!g_{\text{cs}}$, so one can find a number $c_{1}$ satisfying $|\Gcal_{\text{os}}|\!=\!c_{1}g_{\text{cs}}$. Finally, the exponential in Eq.~\eqref{FunctionalOS} vanishes for a purely imaginary gap $\Gcal_{\text{os}}$. Consequently, there exists a number $c$ such that $L_{\text{os}}[q,q^{\prime}]\!=\!c L_{\text{cs}}[q,q^{\prime}]$. Thus, under the conditions stated in Condition~\ref{Cond2}, the Lagrangians are equivalent, and so are the QABs. As an important consequence, one may then compute the brachistochrone from the Hamiltonian dynamics, which presents a comparatively smaller dimension. These specific conditions can be an important tool in scenarios where only partial information about the decoherence is available. 
	
	
	\subsection{STIRAP under balanced loss-gain}
	
	Since the complexity of finding the solution to the QAB depends on the number of independent parameters of the Hamiltonian, the computation of the QAB for a higher-dimensional system can be computed if one identifies a mapping to a lower-dimensional problem. Let us illustrate this point on a three-level system in $\Xi$-configuration, where $\ket{0}$ is the fundamental state and $\ket{1}$ and $\ket{2}$ the excited ones, whose Lagrangian can be mapped into a 2D problem, i.e, with two independent variables $\Omega_{\text{p}}(s)$ and $\Omega_{\text{s}}(s)$.
	Used for STIRAP processes, its Hamiltonian reads~\cite{Vitanov:17}
	\begin{eqnarray}\label{HSTIRAP}
	H_{\text{stirap}}(s) = \hbar \Omega_{\text{p}}(s) \ket{0}\bra{1} + \hbar \Omega_{\text{s}}(s) \ket{1}\bra{2} + \text{h.c.} \text{ , }
	\end{eqnarray}
	where $\Omega_{\text{p}}(0)\!=\!\Omega_{\text{s}}(1)\!=\!0$ and $\Omega_{\text{p}}(1)\!=\!\Omega_{\text{s}}(0)\!=\!\Omega_{0}$. In addition, the system is submitted to a balanced loss-gain decoherence interaction with its reservoir, modelled by the equation~\cite{Dast:14}:
	\begin{eqnarray}\label{RSTIRAP}
	\Rcal_{\text{lg}}[\bullet] = \sum\nolimits_{ n, k } \frac{\Gamma}{2} \left[ 2L^{k}_{n} \bullet L^{k\dagger}_{n} - L^{k\dagger}_{n} L^{k}_{n}\bullet - \bullet L^{k\dagger}_{n} L^{k}_{n} \right] \text{ ,}
	\end{eqnarray}
	where $n\!=\!\{1,2\}$, $k\!=\!\{\text{g},\text{l}\}$, $L^{\text{g}}_{n}\!=\!\ket{n-1}\bra{n}$ and $L^{\text{l}}_{n}\!=\!L^{\text{g}\dagger}_{n}$, with the same transition rate $\Gamma$ for both Lindbladians (balanced case). The eigenvalues of the superoperator are computed using the matrix basis $\{\Sigma_{n} \}$:
	\begin{eqnarray}\label{SigmaStirap}
	\Sigma_{n} = \begin{bmatrix}
	\delta_{1,n} + \delta_{2,n}/\sqrt{3} & \delta_{3,n} + \delta_{1,n} & \delta_{5,n} + \delta_{6,n} \\
	\delta_{3,n} - \delta_{1,n} & -2\delta_{2,n}/\sqrt{3} & \delta_{7,n} + \delta_{8,n} \\
	\delta_{5,n} - \delta_{6,n} & \delta_{7,n} - \delta_{8,n} & \delta_{2,n}/\sqrt{3} -\delta_{1,n}
	\end{bmatrix} \text{ ,}
	\end{eqnarray}
	with $n\!=\!1,\cdots,8$ and $\delta_{1,n}$ the Kronecker delta. The superoperator $\Lmath^{\text{lg}}(s)$ has elements $\Lmath^{\text{lg}}_{kn}(s)\!=\!(1/2)\trs{\Sigma_{k}^{\dagger} \Lcal_{\text{lg}}[ \Sigma_{n} ]}$, and the eigenvalues  $\lambda_{0}\!=\!-3\Gamma/2$, $\lambda_{1}^{\pm}(s)\!=\! [-5\Gamma \pm i \Delta^{(1)}_{\text{rms}}(s)]/4$ (twofold degenerate), and $\lambda_{2}^{\pm}(s)\!=\! -2\Gamma \pm i \Delta^{(2)}_{\text{rms}}(s)$, with $[\Delta_{\text{rms}}^{(n)}(s)]^2\!=\!(16/n^2) \Omega_{\text{rms}}^2(s) - \Gamma^2$ and $\Omega_{\text{rms}}(s)\!=\!\sqrt{\Omega_{\text{s}}^2(s) + \Omega_{\text{p}}^2(s)}$ the root-mean-square Rabi frequency~\cite{Vasilev:09}. We notice that such analytical expressions cannot be obtained for the unbalanced loss-gain configuration, since the characteristic polynomial for $\Lmath^{\text{lg}}(s)$ becomes analytically unsolvable~\cite{Ruffini:Book,Abel:12}.
	
	Using the same criterion to obtain the minimum non-vanishing gap $|\Gcal(s)|$ as the one used for the two-level system, one finds $|\Gcal(s)|\!=\! |\lambda_{2}^{+}-\lambda_{2}^{-}|\!=\!|2i\Delta^{(2)}_{\text{rms}}(s)|$~\footnote{The quantity $|\lambda_{2}^{+}-\lambda_{2}^{-}|$ vanishes if, and only if, $\Gamma\!\rightarrow\!0$ and $\Omega_{0}\!\rightarrow\!0$.}. Up to a constant factor, one obtains the following Lagrangian
	\begin{eqnarray}
	\tilde{L}^{\text{lg}}_{\text{os}}[\Omega(s),\Omega^{\prime}(s)] = \frac{ [\Omega^{\prime}_{\text{s}}(s)]^2 + [\Omega^{\prime}_{\text{p}}(s)]^2 }{\left(  \Omega^2_{\text{p}}(s)+\Omega^2_{\text{s}}(s) -(\Gamma/2)^2 \right)^2}  \text{ . } \label{LthreelsOS}
	\end{eqnarray}
	Hence, the brachistochrone problem for the present three-level system can be mapped to the single-qubit one discussed before, by substituting the decoherence rate $\gamma$ by $\Gamma/2$. Consequently, the same brachistochrone curves are obtained, see Eqs.~\eqref{LtlsOS} and \eqref{LtlsOS2}. It is particularly interesting to observe that under a constant total power constraint ($\Omega^2_{\text{p}}(s)\!+\!\Omega^2_{\text{s}}(s)\!=\!\Omega_{0}^2$), one obtains the pump profiles given in the Eq.~\eqref{LtlsOS2}, which are routinely used in STIRAP protocols~\cite{Vasilev:09,Vitanov:17,Higgins:17,Zhang:20}.  In other words, we have proved that those profiles are the optimal ones for balanced loss-gain systems. 
	Furthermore, this corresponds to the case of equivalent Lagrangians described by Conditions~\ref{Cond2}, for which the QAB does not depend on the decoherence rate. Therefore, such interpolation is the best for both closed and open system case. 
	
\begin{figure}[t!]
		\centering
		\subfloat[ ]{\includegraphics[scale=0.21]{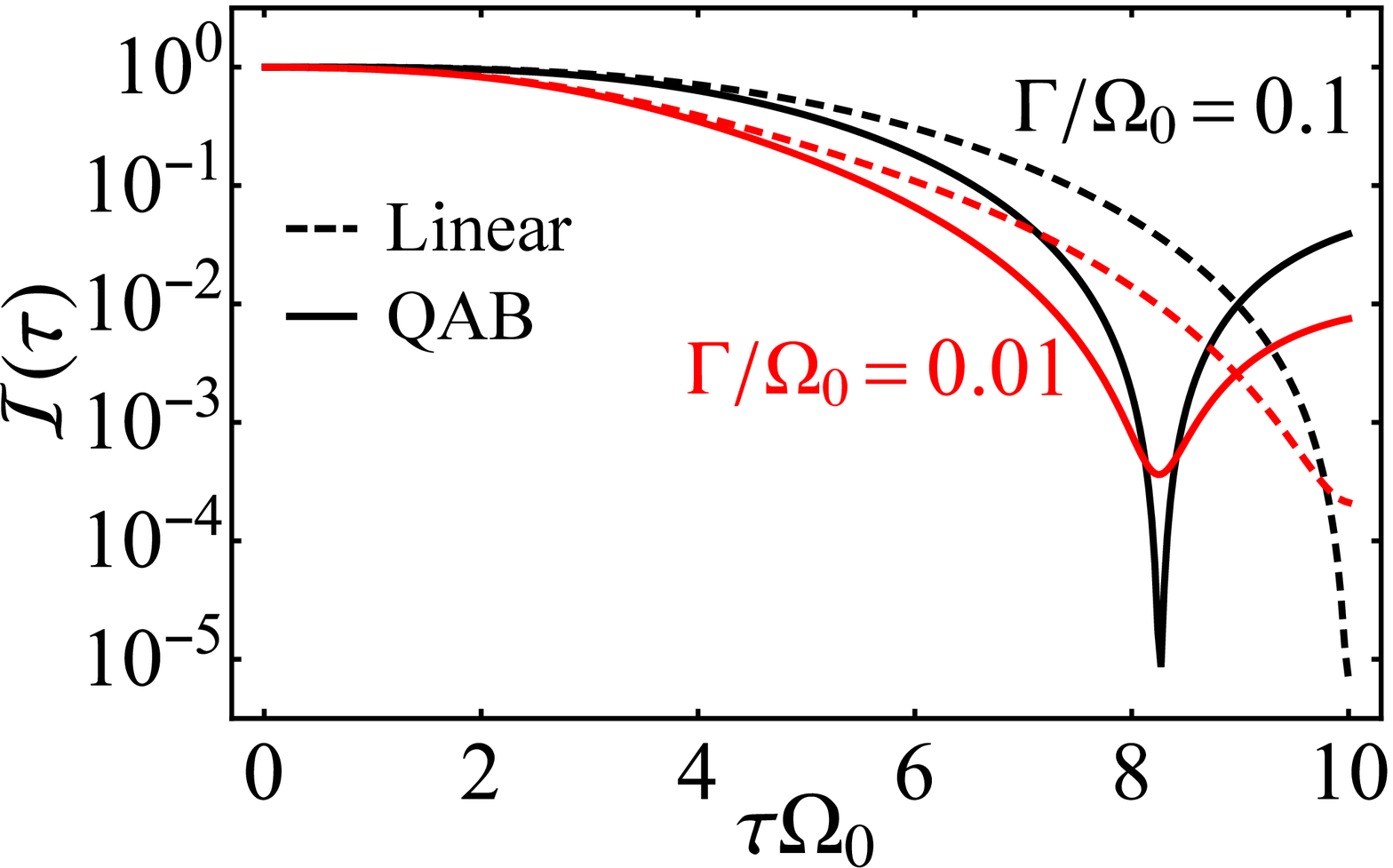}\label{Fig2a}}~
		\subfloat[ ]{\includegraphics[scale=0.2]{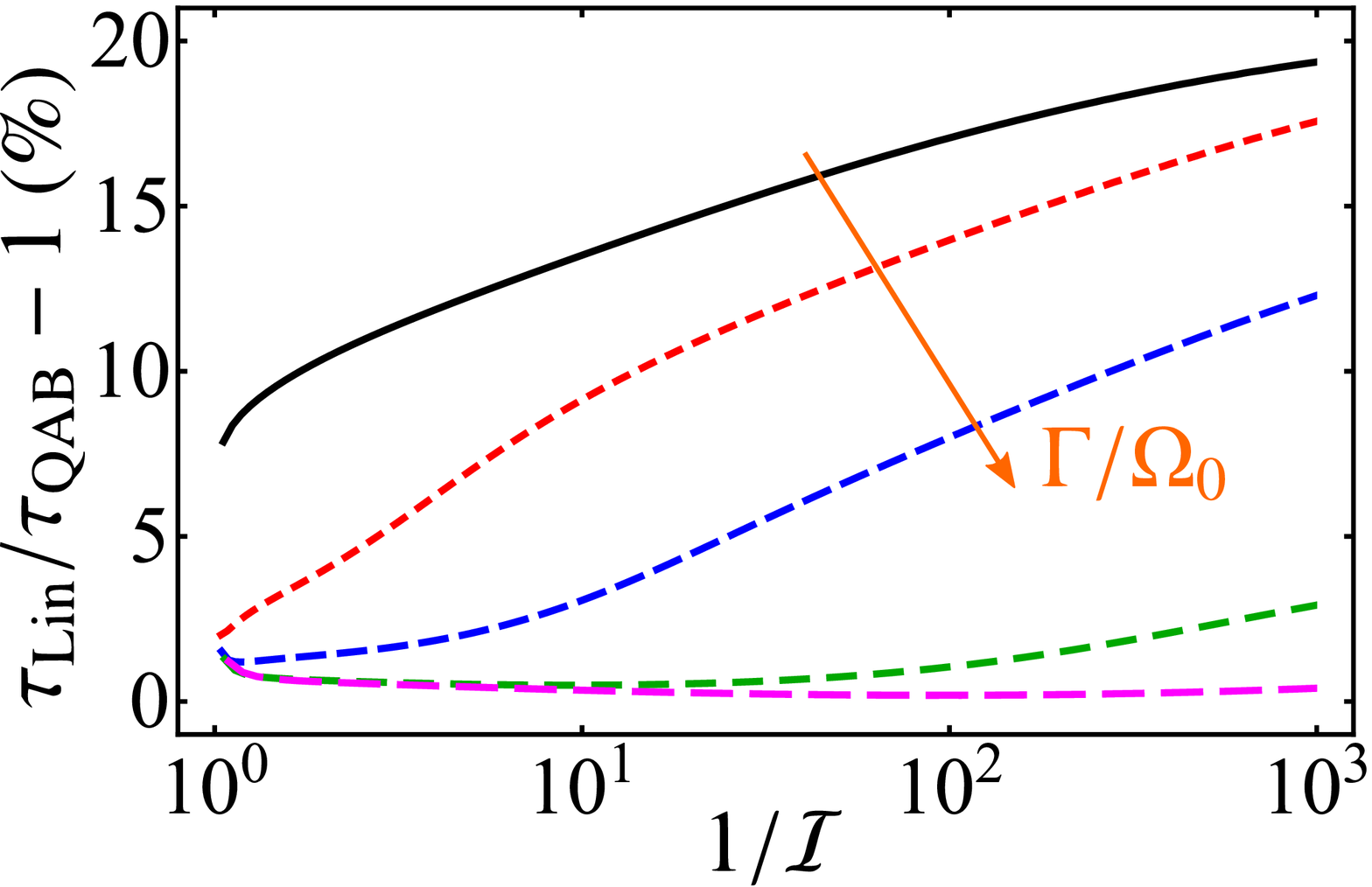}\label{Fig2b}}
		\centering
		\caption{\ref{Fig2a} Infidelity $\Ical$ for the QAB (plain curves) and the linear interpolation $\Omega_{\text{s}}(s)\!=\!\Omega_{0}(1-s)$ (dashed lines), for different decoherence rates [$\Gamma/\Omega_0=0.01$ (red) and $0.1$ (black)]. \ref{Fig2b} Gain $G(\Ical)$ from the brachistochrone over the linear interpolation, as a function of the inverse of the infidelity $1/\Ical$, for different decoherence rates ($\Gamma/\Omega_{0}\!=\!\{0.01, 0.1, 0.2, 0.4\}$ from top to bottom). The arrow denotes the ascending order of values for $\Gamma/\Omega_{0}$.}\label{Fig2}
	\end{figure}

	To quantify the gain of the QAB over another trajectory, it is convenient to introduce the infidelity $\Ical(\tau)\!=\!1-\tr([\rho_f^{1/2}\rho_{\text{ad}}(\tau)\rho_f^{1/2}]^{1/2})$ between the target system state $\rho_f$ and the adiabatic solution $\rho_{\text{ad}}(\tau)$ (see Appendix~\ref{ThreeLevelCase}). The dynamics of the infidelity is presented in Fig.\ref{Fig2}(a), comparing the one for the QAB to that of the linear ramp interpolation: $\Omega_{\text{s}}(s)\!=\!\Omega_{0}(1-s)$. The gain of the QAB can be characterized, for example, by computing the time $\tau$ necessary to reach a given infidelity $\Ical$: The relative gain $G(\Ical)=\tau_\text{Lin}(\Ical)/\tau_\text{QAB}(\Ical)-1$ of the brachistochrone over the linear interpolation is shown in Fig.\ref{Fig2}(b), demonstrating a gain in time of up to $20\%$ to reach an infidelity of $\Ical=10^{-3}$ (under the constraint $\Omega_{\text{p}}(s)+\Omega_{\text{s}}(s)=\Omega_0$). This gain progressively reduces as the dissipation rate is increased, as the dynamics becomes driven by the dissipation rather than by the pump.
	
	\subsection{Adiabatic Deutsch-Jozsa algorithm under dephasing}
	
	Let us now discuss an application to the field of quantum computation: Given a function $f\!:\!x\!\in\!\Rmath\!\rightsquigarrow\!f\!\in\!\Rmath$ promised to be constant or balanced, the Deutsch-Jozsa (DJ) allows us to efficiently determine the nature of $f$~\cite{Deutsch-Jozsa:92}. The adiabatic approach for the DJ algorithm requires an input state $\rho(0)\!=\!\sum_{i=1}^{N} \ket{+_{i}}\bra{+_{i}}$, with $N$ the number of qubits and $\ket{+_{i}}\!=\! (\ket{0}_{i}+\ket{1}_{i})/\sqrt{2}$ the initial state of the $i$-th qubit, so the initial Hamiltonian is $H(0)\!=\!-\hbar\omega\sum_{i=1}^{N} \sigma_{x}^{(i)}$. Since the DJ algorithm features the unitary oracle $\Ocal\!=\!\text{diag}[ (-1)^{f(0)} \cdots (-1)^{f(2^N-1)} ]$, the evolved Hamiltonian reads~\cite{Siu:05} 
	\begin{equation}
	H(s)\!=\!U(s)H(0)U^{\dagger}(s), \text{ with }U(s)\!=\!\exp(i\pi r(s)\Ocal/2),
	\end{equation}
	for some optimal function $r(s)\!\in\![0,1]$ to be identified. In addition, we consider that the qubits are submitted to decoherence, under the form of a local dephasing given by the master equation $\Rcal_{\text{d}}[\bullet]\!=\!\gamma \sum_{k\!=\!1}^{N} [\sigma_{z}^{(k)} \bullet \sigma_{z}^{(k)} - \bullet]$.
	
	Let us now compute the optimal $r(s)$ using the QAB approach: As a first step, we consider the case of $N\!=\!2$ qubits, for which we use the matrix basis $\Sigma\!=\!\{\sigma_{\text{tls}}\}\otimes \{\sigma_{\text{tls}}\}$ to write $\Lmath(s)$. Although we have a $16\times16$-dimensional matrix $\Lmath(s)$, \cvb{its} spectrum 
	can be analytically obtained: It is given by $\lambda_{0}\!=\!0$, $\lambda_{1}\!=\!-2\gamma$, $\lambda_{2}\!=\!-\gamma$ (fourfold degenerate), $\lambda_{3}^{\pm}\!=\!(1/2)(-3\gamma \pm i\bar{\gamma})$ (twofold degenerate), $\lambda_{4}^{\pm}\!=\!(1/2)(-\gamma \pm i\bar{\gamma})$ (twofold degenerate) and $\lambda_{5}^{\pm}\!=\!-\gamma \pm i\bar{\gamma}$, where $\bar{\gamma}^2\!=\!4\omega^2 - \gamma^2$. The time-independent spectrum of $\Lmath(s)$ has a direct impact on the Lagrangian, and $\Gcal$ can now be factorized in the Euler-Lagrange equation. However, due to the exponential in the Lagrangian we need to determine the real part of $\Gcal$; then from the spectrum we obtain the minimum non-vanishing gap $i\gamma$. In what follows, up to a constant factor, the Lagrangian becomes independent on the parameter $\gamma$ and one writes $\tilde{L}^{\text{DJ}}_{\text{os}}[r(s),r^{\prime}(s)]\!=\!r^{\prime 2}(s)$. Moreover, by taking the limit $\gamma\!\rightarrow\!0$, it is straightforward to show that the Lagrangian for the closed case is the same,
	up to a constant factor. Consequently, the Euler-Lagrange equation provides $r^{\prime \prime}(s)\!=\!0$: The solution satisfying $r (0)\!=\!0$ and $r (1)\!=\!1$ provides the optimal interpolation $r (s)\!=\!s$ for both the open and closed system cases. 
	
	The extension of the above result to an arbitrary $N$ is obtained as follows. Firstly, since $H(s)$ derives from a unitary transformation of $H(0)$, the spectrum of $H(s)$ is time-independent, for any $N$. Secondly, for a local dephasing, $\Rcal_{\text{d}}[\bullet]$ and $\Ocal$ are diagonalizable in the same basis, so $\Rcal_{\text{d}}[\bullet]$ only has a time-independent contribution to the spectrum of $\Lmath(s)$. Hence the Lagrangian $\tilde{L}^{\text{DJ}}_{\text{os}}[r(s),r^{\prime}(s)]$ depends only on the contribution of $H^{\prime}(s)$, as in the case of closed systems. Moreover, we find that $H^{\prime}(s)\!=\!i\pi r^{\prime}(s)/2 U(s)[\Ocal,H(0)]U^{\dagger}(s)$, so that the spectrum of $H^{\prime}(s)$ can be written as $h_{n}(s)\!=\! i(\pi/2) b_{n} r^{\prime}(s)$, with $b_{n}$ the (real) eigenvalues of $[\Ocal,H(0)]$. In conclusion, for an arbitrary $N$ we have $\tilde{L}^{\text{DJ}}_{\text{os}}[r(s),r^{\prime}(s)]\!=\!r^{\prime 2}(s)$, up to a constant factor, and the optimal interpolation function is $r(s)\!=\!s$. We stress that despite we obtain a QAB similar to the closed-system one, as a result of satisfying Eq.~\eqref{FunctionalOS}, this property is not even general to the DJ algorithm: This algorithm can be implemented using different time-dependent Hamiltonians~\cite{Saurya:02,Wei:06}, and each particular implementation and each type of decoherence leads to a specific QAB, with different properties.
	
	\section{Conclusion}
	
	We here proposed a generalized approach to find optimal adiabatic interpolation functions through QAB for open systems. The extension from closed to open system is not trivial as it requires computing the Lindbladian spectrum. The variational formalism to find the functional time relies on the notion of an ``adiabatic speed" for open systems, so the optimization procedure can be realized in the frame of adiabatic dynamics. The examples provided allowed to demonstrate the potential of the method for quantum information schemes, so it could find a direct application for trapped ions~\cite{Hu:18,Hu:19-a} and nuclear magnetic resonance systems~\cite{Santos:20b}, for example.
	
	Note that the QAB discussed up to now have analytical forms, which can typically be achieved only for small or specific systems such as the DJ algorithm. However, due to the generality of our approach, the theory developed here can be applied to any physical system driven by Eq.~\eqref{DynSuper} where exact simulations are possible. More generally, the QAB for an arbitrary open system can be obtained numerically by integrating the second-order differential equations provided by the Euler-Lagrange equation, see Appendix~\ref{NumericalSec} for more details on the numerical procedure. Such numerical computation suffers the same limitation as any exact simulation: it requires the diagonalization of the superoperator (and its derivative), thus limiting the numerical approach to $\!\sim\!10$ two-level systems for open systems and $\!\sim\!20$ ones for closed systems. In this context, approximated methods may represent a promising path to explore the QAB for larger systems~\cite{Horner:19,Antoulas:04,Rotureau:06,Rotureau:09}, with potential applications to quantum thermodynamics~\cite{Hu:20-a,Dann:20}, open quantum many-body systems~\cite{Patane:09}, among others~\cite{Huang:08,Jing:13,Thunstrom:05}.
	
	\begin{acknowledgments}
		A. C. S., C.J.V.-B. and R.B. acknowledge the financial support of the São Paulo Research Foundation (FAPESP) (grant 2018/15554-5, 2019/22685-1, and 2019/11999-5) and  the Coordenação de Aperfeiçoamento de Pessoalde Nível Superior (CAPES/STINT), grant 88881.304807/2018-01. R.B. and C.J.V.-B. benefited from the support of the National Council for Scientific and Technological Development (CNPq) Grant Nos. 302981/2017-9, 409946/2018-4, and 307077/2018-7. C.J.V.-B. also thanks the support from the Brazilian National Institute  of  Science  and  Technology  for  Quantum Information (INCT-IQ/CNPq) Grant No. 465469/2014-0. 
	\end{acknowledgments}
	
	\appendix
	
	\section{Derivation of the Eq.~(1)} \label{derivEq1}
	
	Let us consider the system dynamics given by the master equation
	\begin{align}
	\dot{\rho}(t) = \Lcal[\rho(t)] , \label{EqEqLind}
	\end{align}
	where $\Lcal[\bullet]$ is the dynamics generator. We can deal with above dynamics in different ways, but here we will consider the case in which we define a extended space where the superoperator becomes an $(D^2 \times D^2)$-dimensional operator $\Lmath(t)$, where $D$ is the dimension of the Hilbert space, and the operators (as density matrix and observables, for example) becomes vectors $\dket{\rho(t)}$ on such extended space. To see how it can be done, we start designing a set of $D^2 -1$ operators basis $\Ocal = \{\sigma_{n}\} \in \Hcal_{\text{S}}$, so that $\trs{\sigma_{n}} = 0$ and $\trs{\sigma_{n}\sigma_{m}^{\dagger}} = D\delta_{nm}$, $\rho(t)$ can be written as
	\begin{eqnarray}
	\rho(t) = \frac{1}{D} \left[ \1 + \sum _{n=1}^{D^2 -1} \varrho_{n}(t) \sigma_{n} \right] \text{ , } \label{EqEqRhoCoherence}
	\end{eqnarray}
	where the identity is introduced in this way in order to satisfy the property $\trs{\rho(t)} = 1$ and the coefficients $\varrho_{n}(t) = \trs{\rho(t)\sigma_{n}^{\dagger}}$. Thus, if we substitute $\rho(t)$ in Eq.~\eqref{EqEqLind} by using the expanded form of Eq.~\eqref{EqEqRhoCoherence}, we find the system of differential equations
	\begin{eqnarray}
	\dot{\varrho}_{k} (t) = \frac{1}{D} \sum_{n=0}^{D^2-1} \varrho_{i}(t) \trs{\sigma_{k}^{\dagger} \Lcal [ \sigma_{i} ]} \text{ . } \label{ApEqv1}
	\end{eqnarray}
	where we assume that $\Lcal [\bullet]$ is a linear superoperator and denote $\sigma_{0} = \1$. Note that if we identify the coefficient $\trs{\sigma_{k}^{\dagger} \Lcal [ \sigma_{i} ]}$ in above equation as am element at $k$-th row and $i$-th column of a $D^2 \times D^2$-dimensional matrix $\Lmath(t)$, one can write
	\begin{eqnarray}
	\dket{\dot{\rho}(t)} = \Lmath(t) \dket{\rho(t)} \text{ , } \label{ApEqSuperLindEq}
	\end{eqnarray}
	where $\dket{\rho(t)}$ is a $D^2$-dimensional vector with components $\varrho_{n}(t) = \trs{\rho(t)\sigma_{n}^{\dagger}}$, $n=0,1,\cdots D^2-1$.
	
	\section{Adiabatic dynamics for the three-level system} \label{ThreeLevelCase}
	
	We consider the evolution of a three-level driven by the master equation given by $\Lcal_{\text{lg}}[\bullet]\!=\!(1/i\hbar)[H(t),\bullet] + \Rcal_{\text{lg}}[\bullet]$, where $H_{\text{stirap}}(s)$ and $\Rcal_{\text{lg}}[\bullet]$ are given by the Eqs.~({\color{blue}10}) and~({\color{blue}11}) of the main text. To study the adiabatic dynamics of the system, let us rewrite the master equation $\Lcal_{\text{lg}}[\bullet]$ in the superoperator formalism. In order to simplify the description of the three-level system dynamics in open system, we can choose a matrix basis in which the density matrix reads
	\begin{align}
	\rho(t) = \frac{1}{D} \1 + \frac{1}{d}\sum_{n=1}^{8} \rho_{n}(t) \sigma_{n} ,
	\end{align}
	where the first term guarantees that $\trs{\rho(t)}\!=\!1$ and the coefficients are given by $\rho_{n}(t)\!=\!\trs{\rho(t)\sigma_{n}^{\dagger}}$, $d$ is a dimensionless parameter. Now, by using the above expression in the dynamical equation we get
	
	\begin{align}
	\dot{\rho}(t) = \frac{1}{d}\sum_{n=1}^{8} \dot{\rho}_{n}(t) \sigma_{n} = \Lcal_{\text{lg}}[\rho(t)] \text{ , }
	\end{align}
	so that the dynamics of the component $\rho_{k}(t)$ can be obtained as
	\begin{align}
	\trs{\dot{\rho}(t)\sigma_{k}^{\dagger}} &= \frac{1}{d}\sum_{n=1}^{8} \dot{\rho}_{n}(t) \trs{\sigma_{n}\sigma_{k}^{\dagger}} = \trs{\Lcal_{\text{lg}}[\rho(t)]\sigma_{k}^{\dagger}} \nonumber \\
	\dot{\rho}_{n}(t) &= \frac{1}{D}\trs{\Lcal_{\text{lg}}[\1]\sigma_{k}^{\dagger}} + \frac{1}{d}\sum_{n=1}^{8}\rho_{n}(t) \trs{\Lcal_{\text{lg}}[\sigma_{n}]\sigma_{k}^{\dagger}} \nonumber \\&= \frac{1}{d}\sum_{n=1}^{8}\rho_{n}(t) \trs{\Lcal_{\text{lg}}[\sigma_{n}]\sigma_{k}^{\dagger}} , \label{ApSetEq}
	\end{align}
	where we have used that $\Lmath^{\text{lg}}_{kn}(t)\!=\!(1/d)\trs{\sigma_{k}^{\dagger}\Lcal_{\text{lg}}[\sigma_{n}]}$ and $\Lcal_{\text{lg}}[\1]\!=\!0$. It is then clear that the component $\1$ of the density matrix does not evolve in time, so that our analysis can be reduced to that of a $8\!\times\!8$ superoperator $\Lmath_{\text{lg}}(t)$. In conclusion, we can write the set of equations given in Eq.~\eqref{ApSetEq} as
	\begin{align}
	\dket{\dot{\rho}(t)} = \Lmath_{\text{lg}}(t)\dket{\rho(t)} ,
	\end{align}
	with the components of $\dket{\rho(t)}$ given by $\rho_{k}(t)$. Now, we define the set of matrix $\{\sigma_{j}\}$ as given by the Eq.~({\color{blue}12}) in the main text, which satisfies $\trs{\sigma_{j}\sigma_{n}^{\dagger}}\!=\!2\delta_{nj}$. In this basis the superoperator $\Lcal_{\text{lg}}[\bullet]$ is given by 
	\begin{widetext}
		\begin{align}\label{APLmath}
		\Lmath_{\text{lg}}(s)=\begin{bmatrix}
		-\Gamma & 0 & 0 & i\Omega_{\text{p}}(s) & 0 & 0 & 0 & i\Omega_{\text{s}}(s) \\
		0 & -3\Gamma  & 0 & i\sqrt{3}\Omega_{\text{p}}(s) & 0 & 0 & 0 & -i\sqrt{3}\Omega_{\text{s}}(s) \\
		0 & 0 & -3\Gamma/2 & 0 & 0 & i\Omega_{\text{s}}(s) & 0 & 0 \\
		i \Omega_{\text{p}}(s) & i\sqrt{3} \Omega_{\text{p}}(s) & 0 & -3\Gamma/2 & i \Omega_{\text{s}}(s) & 0 & 0 & 0 \\
		0 & 0 & 0 & i \Omega_{\text{s}}(s) & -\Gamma & 0 & 0 & -i\Omega_{\text{p}}(s) \\
		0 & 0 & i \Omega_{\text{s}}(s) & 0 & 0 & -\Gamma & -i\Omega_{\text{p}}(s) & 0 \\
		0 & 0 & 0 & 0 & 0 & -i\Omega_{\text{p}}(s) & -3\Gamma/2 & 0 \\
		i\Omega_{\text{s}}(s) & -i\sqrt{3}\Omega_{\text{s}}(s) & 0 & 0 & -i\Omega_{\text{p}}(s) & 0 & 0 & -3\Gamma/2
		\end{bmatrix} ,
		\end{align}
	\end{widetext}
	with the set of eigenvalues of $\Lmath_{\text{lg}}(s)$ shown in the main text. Let us now study the adiabatic dynamics of the system, with the initial state
	\begin{align}
	\rho(0) = \frac{1}{D} \1 + \frac{1}{d}\sum_{n=1}^{8} \rho_{n}(0) \sigma_{n} ,
	\end{align}
	which defines the initial ``coherence vector" $\dket{\rho(0)}$ with components $\rho_{n}(0)$. In particular, by considering the matrix notation for the states as $\ket{n}\!=\![\delta_{0n}~\delta_{1n}~\delta_{2n}]^{t}$ (the superscript ``t" refers to the transpose) and the system starting in state $\ket{0}$, we get
	\begin{align}
	\dket{\rho(0)} = \begin{bmatrix} 1 & 1/\sqrt{3} & 0 & 0 & 0 & 0 & 0 & 0\end{bmatrix}^{t} . \label{Aprho0}
	\end{align}
	Then, by using the solution for the adiabatic dynamics in open system we write ($\tau$ being the total evolution time)
	\begin{align}
	\dket{\rho_{\text{ad}}(s)} = \sum_{n} c_{n} \exp\left(\tau \int_{0}^{s}\Theta_{n}(\xi)d\xi\right) \dket{\Dcal_{n}(s)} , 
	\end{align}
	where $c_{n}$ are constants computed from the initial state $\dket{\rho(0)}$, $\dket{\Dcal_{n}(s)}$ the right eigenvector of $\Lmath_{\text{lg}}(s)$ with eigenvalue $\lambda_{n}(s)$ ($\dbra{\Ecal_{n}(s)}$ being the left one), and $\Theta_{n}(s)\!=\!\lambda_{n}(s)-(1/\tau)\dinterpro{\Ecal_{n}(s)}{\Dcal^{\prime}_{n}(s)}$ the instantaneous adiabatic quantal phases of the evolution. 
	Remark that the number of right and left eigenvectors required to determine $\dket{\rho_{\text{ad}}(s)}$ depends on the number of non-vanishing parameters $c_{n}$. Diagonalizing the matrix $\Lmath_{\text{lg}}$,  we conclude that the relevant eigenvectors are
	\begin{subequations}\label{APEigen}
		\begin{align}
		\dket{\Dcal_{0}(s)} &= \frac{2i \Omega_{0} f_{-}(s)}{\Gamma}\dket{1} - \frac{2i\Omega_{0} f_{+}(s)}{\sqrt{3}\Gamma}\dket{2} - f_{\text{p}}(s) \dket{4}  \\ &+ \frac{4i\Omega_{0} f_{\text{p}}(s)}{\Gamma}\dket{5} + f_{\text{s}}(s)\dket{8} , \nonumber \\
		\dket{\Dcal_{-}(s)} &= \frac{i\Omega_{0}f_{-}(s)}{\Gamma_{+}(s)}\dket{1} -  \frac{i \sqrt{3}\Omega_{0} \Gamma_{+}(s)}{4\Omega_{0}}\dket{2} - f_{\text{p}}(s) \dket{4} \nonumber \\ &+ \frac{2i \Omega_{0}f_{\text{p}}(s)f_{\text{s}}(s)}{\Gamma_{+}(s)}\dket{5} + f_{\text{s}}(s)\dket{8} ,\\
		\dket{\Dcal_{+}(s)} &= -\frac{i\Omega_{0}f_{-}(s)}{\Gamma_{-}(s)} \dket{1} - \frac{i \sqrt{3}\Gamma_{-}(s)}{4\Omega_{0}} \dket{2} - f_{\text{p}}(s)\dket{4} \nonumber \\ &+ \frac{2i\Omega_{0} f_{\text{p}}(s)f_{\text{s}}(s)}{\Gamma_{-}(s)}\dket{5} + f_{\text{s}}(s)\dket{8}
		, 
		\end{align}
	\end{subequations}
	with eigenvalues $\lambda_{0}$ and $\lambda_{2}^{\pm}(s)$ as defined in main text. We defined dimensionless functions $f_{\text{p}/\text{s}}(s)$ from $\Omega_{\text{p}/\text{s}}(s)\!=\!\Omega_{0}f_{\text{p}/\text{s}}(s)$, $\Gamma_{\pm}(s)\!=\!\Gamma \pm \sqrt{\Gamma^2-4\Omega_{0}f_{+}(s)}$, with $f_{+}(s)\!=\!f_{\text{p}}(s)-f_{\text{s}}(s)$, and $\dket{n}=[\delta_{1n}~\delta_{2n}~\cdots~\delta_{8n}]$ the coherence vector canonical basis. From the Eqs.~\eqref{Aprho0} and~\eqref{APEigen} we determine the non-vanishing constants $c_{n}$ from the initial state:
	\begin{align}
	c_{0} &= \frac{8i\Gamma\Omega_{0}}{16\Omega_{0}^{2}-3\Gamma^2} , ~ c_{\pm} = -\frac{2i\Gamma\Omega_{0}}{16\Omega_{0}^{2}-3\Gamma^2}\left(2\pm \frac{\Gamma}{\sqrt{\Gamma^2-4\Omega_{0}^{2}}}\right) .
	\end{align}
	Thus, the evolved state reads
	\begin{align}
	\dket{\rho_{\text{ad}}(s)} = \sum_{n=\{0,\pm\}}c_{n} e^{\tau \int_{0}^{s} \lambda_{n}(\xi) - (1/\tau)\vartheta_{n}(\xi)d\xi} \dket{\Dcal_{n}(s)} , 
	\end{align}
	with
	\begin{align}
	\vartheta_{0}(s) &= \frac{f^{\prime}_{+}(s)\left[ 3\Gamma^2 - 32 \Omega_{0}^{2}f_{+}(s) \right]}{2f_{+}(s)\left[ 16 \Omega_{0}^{2}f_{+}(s) -3\Gamma^2 \right]} , \\
	\vartheta_{\pm}(s) &= \Omega_{0}^{2} f^{\prime}_{+}(s) \frac{2\Gamma \sqrt{\Gamma^2 - 4 \Omega_{0}^{2}f_{+}(s) }\mp\left[32\Omega_{0}^{2}f_{+}(s) - 7\Gamma^2\right]}{\left[\Gamma^2 - 4 \Omega_{0}^{2}f_{+}(s)\right] \left[ 3\Gamma^2 - 16 \Omega_{0}^{2}f_{+}(s) \right]} .
	\end{align}
	
	Finally, using the coherence vector in the basis $\dket{n}$, we can recover the density matrix $\rho(t)$ of the system by applying the inverse transformation from superoperator to operator description as
	\begin{align}
	\rho_{\text{ad}}(s) = \frac{1}{D} \1 + \frac{1}{d}\sum_{n=1}^{8} \dinterpro{n}{\rho_{\text{ad}}(s)} \sigma_{n} .
	\end{align}

	\section{Numerical computation of quantum adiabatic brachistochrone for open systems} \label{NumericalSec}
	
	We here present the procedure to compute numerically the brachistochrone for an arbitrary open and closed systems, the limit of which relies in the system size which can be handled numerically, as we shall see.
	Calling $q_n$ the components of the vector $q(t)$, the Euler-Lagrange equation leads to the following set of differential equations:
	\begin{align}
	\frac{d}{dt}\left[\frac{\partial L(q,\dot{q})}{\partial \dot{q}_{n}}\right] - \frac{\partial L(q,\dot{q})}{\partial q_{n}} = 0.\label{eq:EL}
	\end{align}
	
	The Lagrangian in our approach decomposes as $L(q,\dot{q}) = Q(q) V(\dot{q},q)$,
	where $Q(q)$ accounts for the gap structure and $V(\dot{q},q)$ is the squared norm of the Lindbladian superoperator derivative:
	\begin{align}\label{QV}
	V(\dot{q},q) &= ||\dot{\Lmath}(t)||^2,\\
	Q(q) &= \left[\frac{1}{|\Gcal(t)|^2}\exp(\int_{0}^{t}\Re[\Gcal(\xi)]d\xi)\right]^2.
	\end{align}
	This allows us to rewrite the Euler-Lagrange equation \eqref{eq:EL} as
	\begin{align}
	\frac{d}{dt}\left[\frac{\partial L(q,\dot{q})}{\partial \dot{q}_{n}}\right] &= \frac{d}{dt}\left[\frac{\partial [Q(q) V(\dot{q},q)]}{\partial \dot{q}_{n}}\right] \nonumber \\
	&=\sum_{k} \left[\frac{\partial Q(q)}{\partial q_{k}} \dot{q}_{k} \frac{\partial V(\dot{q},q)}{\partial \dot{q}_{n}} + Q(q) \frac{\partial^2 V(\dot{q},q)}{\partial\dot{q}_{k}\partial \dot{q}_{n}} \ddot{q}_{k}\right. \nonumber \\
	&+\left. Q(q) \frac{\partial^2 V(\dot{q},q)}{\partial q_{k}\partial \dot{q}_{n}} \dot{q}_{k} \right] .
	\end{align}
	
	
	Then, from Eq.~\eqref{eq:EL}, we find
	\begin{align}
	\sum_{k}& \left[ Q(q) \frac{\partial^2 V(\dot{q},q)}{\partial\dot{q}_{k}\partial \dot{q}_{n}} \ddot{q}_{k} + \left(\frac{\partial Q(q)}{\partial q_{k}} \frac{\partial V(\dot{q},q)}{\partial \dot{q}_{n}} + Q(q) \frac{\partial^2 V(\dot{q},q)}{\partial q_{k}\partial \dot{q}_{n}}\right) \dot{q}_{k}\right] \nonumber \\
	& - \frac{\partial [Q(q) V(\dot{q},q)]}{\partial q_{n}} = 0 . \label{edo1}
	\end{align}
	
	Thus, the problem of computing the brachistochrone reduces to a set of coupled second-order differential equation. Although their number may seem reduced (one for each independent variable of the driving field), it does not represent the true complexity of this calculation. Indeed at each time step the gap $\Gcal(t)$ and the norm of the superoperator derivative $\dot{\Lmath}(t)$ must be computed, so the computation of the brachistochrone faces the same limitation as any exact simulation of arbitrary open systems: it requires computing the spectrum of a matrix of side $D^2$, for open systems, and side $D$ for closed systems, with $D$ the dimension of the Hilbert space. Hence, it is typically limited to open systems of up to around ten qubits.
	
	Finally, let us comment on some aspects of this computation. First, as any brachistochrone problem, the boundary conditions correspond to initial and final states, $q(0)$ and $q(1)$, and the derivatives $q'(0)$ and $q'(1)$ are not set. In this sense, the brachistochrone problem presents Dirichlet boundary conditions, which suggests the use of a dedicated solver. Second, there are several characteristics of the problem which may significantly simplify the computation of the brachistochrone: the existence of constraints (i.e., a relation between the different components of the driving $q$), the possible time-independent nature of the Hamiltonian and Lindbladian (as in the examples of the main text) or the simple dependence of the superoperator on the driving fields (a common physical situation is a Hamiltonian which depends linearly on these fields) are such examples. We shall now illustrate these points with a three-level system for which the brachistochrone problem does not possess an analytical solution.
	
	
	As an example of numerical computation of the brachistochrone, we study a single transmon qutrit. The non-unitary dynamics of this three-level system encoded in the energy levels of a superconducting transmon is given by~\cite{Peterer:15}
	\begin{eqnarray}
	\dot{\rho}(t) = \frac{1}{i\hbar} [H(t),\rho(t)] + \Lcal_{\text{rel}}[\rho(t)]+ \Lcal_{\text{dep}}[\rho(t)] \text{ , } \label{LindEq}
	\end{eqnarray}
	where $\Lcal_{\text{rel}}[\bullet]$ and $\Lcal_{\text{dep}}[\bullet]$ describe the relaxation and dephasing phenomena, respectively:
	\begin{subequations}
		\begin{align}
		\Lcal_{\text{rel}}[\bullet] &= \sum_{k\neq j}\Gamma_{kj} \left[\sigma_{kj}\bullet\sigma_{jk} - \frac{1}{2}\{\sigma_{kk},\bullet\} \right] \text{ , } \\
		\Lcal_{\text{dep}}[\bullet] &= \sum_{j=2,3}\gamma_{j} \left[\sigma_{jj}\bullet\sigma_{jj} - \frac{1}{2}\{\sigma_{jj},\bullet\} \right].
		\end{align}
		\label{RelTerm}
	\end{subequations} 
	with $\sigma_{kj}=\ket{k}\bra{j}$ and $\Gamma_{kj}\!=\!\Gamma_{jk}$. The Hamiltonian for the problem is given as in the Eq.~({\color{blue}10}) of the main text, with $\Omega_{\text{p}}(s)\!=\!\Omega_{0} q_{1}(s)$ and $\Omega_{\text{s}}(s)\!=\!\Omega_{0} q_{2}(s)$, with $s\!=\!t/\tau$. 
	
	Because of the linear dependency of $H$ on the driving fields $q$, $V$ is naturally a quadratic function of $q'$. If we consider in addition the constraint $\Omega_{\text{p}}(s)+\Omega_{\text{s}}(s)=\Omega_0$, we obtain $q^{\prime}_{1}(s)\!=\!- q^{\prime}_{2}(s)\equiv q'(s)$ and then $V=\alpha q^{\prime 2}(s)$, with $\alpha$ a constant (more specifically, writing $\Lmath(s)\!=\! q^{\prime}_{2}(s) \Lmath_{0}$, we obtain $\alpha=||\Lmath_{0}||^2$).
	
	\begin{figure*}[t!]
		\subfloat[ ]{\includegraphics[scale=0.472]{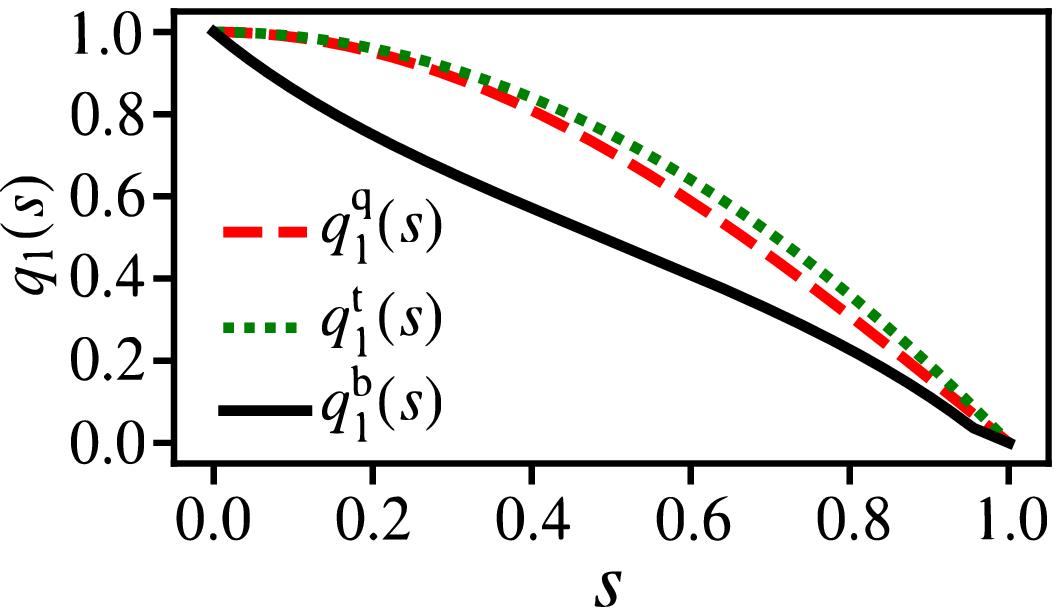}\label{FigSol}}\quad
		\subfloat[ ]{\includegraphics[scale=0.472]{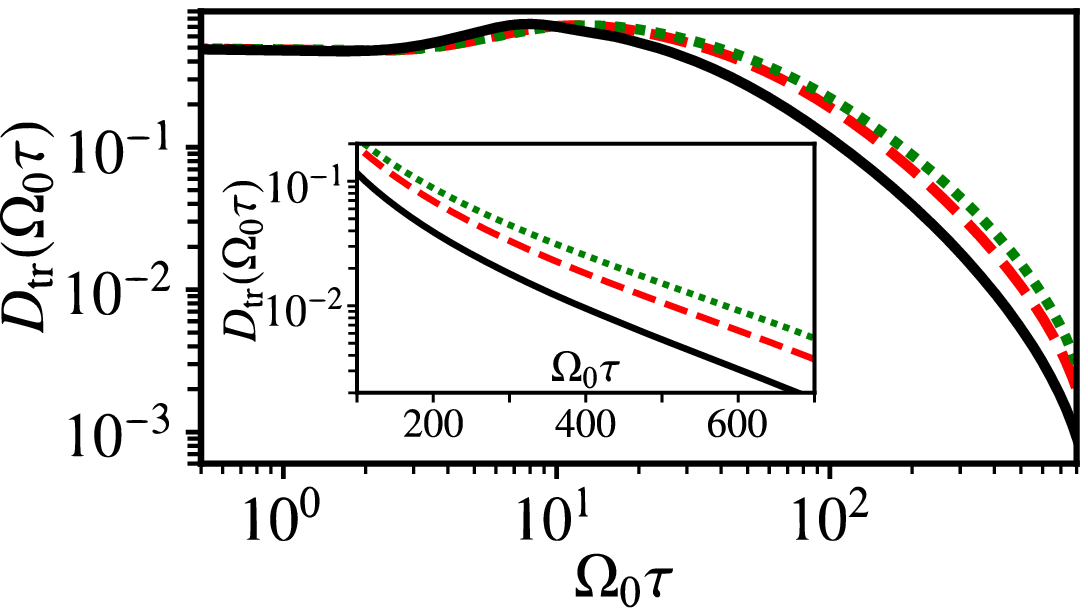}\label{FigTD}}\quad
		\subfloat[ ]{\includegraphics[scale=0.472]{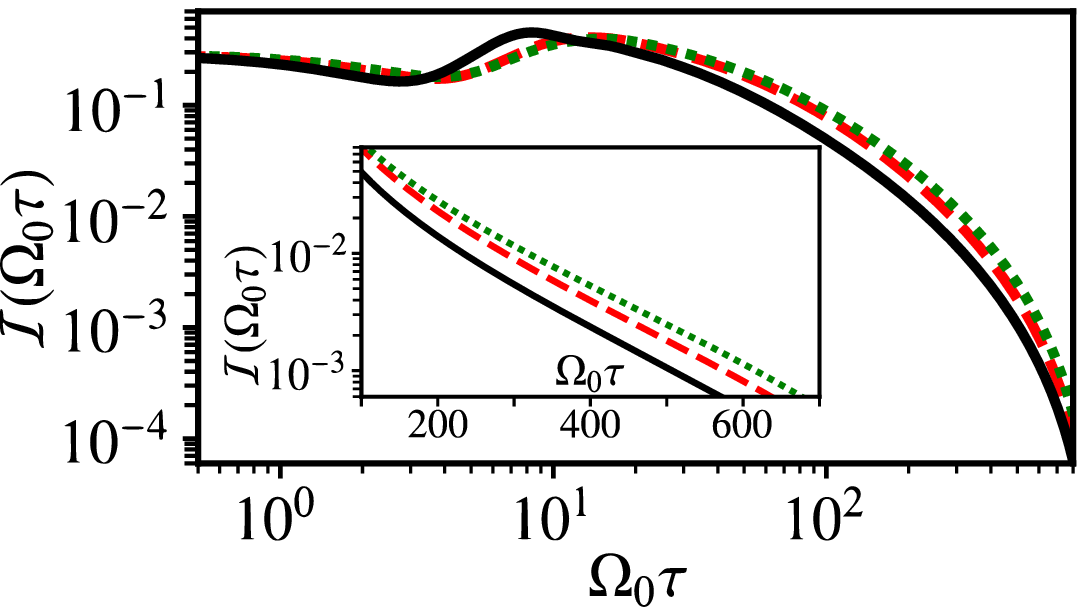}\label{FigInf}}
		\caption{(a) Driving field corresponding to the Brachistochrone curve $q_{1}^{\text{b}}(s)$ (black plain curve), to a quadractic interpolation, $q_{1}^{\text{q}}(s)$ (red dashed), and to a cosine one, $q_{1}^{\text{t}}(s)$. (b) Fidelity to the adiabatic solution, given by the trace distance \eqref{Dist}, and (c) infidelity for each dynamics. The inset figures in (b) (trace distance) and in (c) (infidelity) highlight the gain provided by the brachistochrone.}
		\label{FigNumSol}
	\end{figure*}
	
	
	In this particular case of a single independent component of the driving field, the brachistochrone differential equation \eqref{edo1} turns into
	\begin{align}
	q^{\prime\prime}+\frac{q^{\prime 2}}{2  Q(q)} \frac{\partial Q(q)}{\partial q} = 0.\label{Case2}
	\end{align}

	The differential equation \eqref{Case2} is solved numerically using a Python routine, and the resulting brachistochrone for the driving field, $q_1^\text{b}(s)$ is presented in Fig.~\ref{FigNumSol}(a). As an example, we compare it to quadratic, $q_1^\text{q}(s)=1-s^2$, and cosine-like, $q_1^\text{t}(s)=1-\cos(\pi s/2)$, interpolations.
	In order to illustrate the gain from the brachistochrone, we compare the fidelities from the different trajectories using the trace distance
	\begin{align}
	D_{\text{tr}}(\rho_{1},\rho_{2}) &= \frac{1}{2}\text{Tr}\left[ \sqrt{\left(\rho_{1}-\rho_{2}\right)^{\dagger}\left(\rho_{1}-\rho_{2}\right)}\right] \label{Dist}
	\end{align}
	and the infidelity (see main text). As shown in Fig.~\ref{FigNumSol}(b) and (c), the brachistochrone brings the system close to its target state in a more efficient way, as expected.
	
	
	
	%

\end{document}